\documentstyle[prb,aps,epsf]{revtex}
\widetext
\begin{document}
\draft
\title{The Heisenberg antiferromagnet on an anisotropic triangular lattice:
linear spin-wave theory}
\author{J. Merino\cite{email} and Ross H. McKenzie}
\address{School of Physics, University of New South Wales, Sydney 2052, Australia}
\author{J. B. Marston and C. H. Chung}
\address{Department of Physics, Brown University, Providence, RI-02912-1843}
\bigskip
\maketitle

\begin{abstract}
We consider the effect of quantum spin fluctuations on the ground 
state properties of the Heisenberg
antiferromagnet on an anisotropic triangular lattice using linear spin-wave
(LSW) theory. This model should describe the magnetic properties of the
insulating phase of the $\kappa-(BEDT-TTF)_2 X$ family of superconducting
molecular crystals. The ground state energy, the staggered magnetization, magnon
excitation spectra and spin-wave velocities  are computed
as a function of the ratio of the antiferromagnetic exchange between the 
second and first neighbours, $J_2/J_1$.
We find that near $J_2/J_1 = 0.5 $, {\it i.e.}, in the region where
the classical spin configuration changes from a N\'eel ordered phase 
to a spiral phase, the staggered magnetization vanishes, suggesting 
the possibility of a quantum disordered state. 
In this region, the quantum correction to the magnetization is large 
but finite. This is in contrast to the frustrated Heisenberg model 
on a square lattice, for which the quantum correction diverges
logarithmically at the transition from the N\'eel to the collinear 
phase. For large $J_2/J_1$, the model becomes a set of chains 
with frustrated interchain coupling. For $J_2 > 4 J_1$, the quantum   
correction to the magnetization, within LSW theory, becomes comparable
to the classical magnetization, suggesting the possibility of a quantum
disordered state. We show that, in this regime, the quantum fluctuations
are much larger than for a set of weakly coupled chains with non-frustated
interchain coupling. 
\end{abstract}


\bigskip
\section{Introduction}
\label{sec:intro}
The study of strongly correlated electron systems in low dimensions is
a very active field of research. One of the great challenges is to understand
the competition between antiferromagnetism and superconductivity found
in cuprate and organic superconductors. 
Kino and Fukuyama\cite{Kino:96}recently proposed interacting electron models for a 
range of BEDT-TTF crystals. McKenzie argued that the $\kappa-(BEDT-TTF)_2X$
family can be described by a simplified version of one of their models, a 
Hubbard model on an anisotropic triangular lattice\cite{McKenzie:98}. Recent
Quantum Monte Carlo calculations\cite{Kuroki:98} and 
calculations at the level of the random-phase approximation\cite{Votja:98}
and the fluctuation-exchange approximation \cite{Schmalian:98}
suggest that at the boundary of the antiferromagnetic phase,
this model exhibits superconductivity mediated by spin fluctuations. As 
the anisotropy of the intersite hopping varies, the model changes from the
square lattice to the isotropic triangular lattice to decoupled chains
\cite{McKenzie:98}. The
wavevector associated with the antiferromagnetic spin fluctuations changes and 
the superconductivity has been predicted to change from d-wave singlet (as
in the cuprates) to s-wave triplet in the odd-frequency channel for the
isotropic triangular lattice \cite{Votja:98}.
 This shows that an understanding of the 
antiferromagnetic interactions is important for understanding the symmetry 
of the Cooper pairs in the superconducting state.     
 As suggested by Seo and Fukuyama\cite{Seo:98}, $\theta-(BEDT-TTF)_2RBZn(SCN)_4$ 
can be described by the model we consider with $J_2/J_1 \approx 5 $.
Experimental findings from Mori {\it et al.}\cite{Mori:97}
show that this material has a spin-gap.
The same model has also been proposed by Horsch and Mack\cite{Horsch:98} to
be relevant to $\alpha$\'-$NaVO_5$.

The Heisenberg model studied here 
should also describe the magnetic properties of the molecular crystals  
$\kappa-(BEDT-TTF)_2X$ with $X=Cu[N(CN)_2]Cl$, $Cu(CN)_3$ and $d_8-Cu[N(CN)_2]Br$
\cite{note}
which are non-metallic at ambient pressure\cite{Komatsu:96,Kanoda:97}. On
the basis of NMR lineshapes, Kanoda\cite{Kanoda:97} has suggested that the magnetic 
ordering in $\kappa-(BEDT-TTF)_2Cu[N(CN)_2]Cl$ and $d_8-Cu[N(CN)_2]Br$ 
are commensurate.
The magnetic moment has been estimated to be (0.4-1.0) $\mu_B$ per dimer.
Using uniaxial stress within a layer or changing the anion $X$, it may
be possible to vary the ratio $J_2/J_1$ and induce a quantum phase 
transition into a disordered phase or the spiral phase discussed  
here. 

This paper is organized as follows: in Section \ref{sec:mod}, we introduce
the model, before presenting linear-spin wave theory in Section \ref{sec:theo}.
In Section \ref{sec:res} we present our results.  We find that at
$J_2/J_1=0.5$ quantum fluctuations are enhanced giving a large 
but finite correction to the magnetization, suggesting the possibility of
having a disordered state. This possibility is also found   
for $J_2/J_1 > 4$: in this region of parameters, the quantum correction to 
the magnetization is comparable to the classical magnetization and
we find that is much larger than for a set of weakly coupled chains with
non-frustrated interchain couplings.

\section{The Model}
\label{sec:mod}
\begin{figure}
\center
\epsfxsize = 10cm \epsfbox{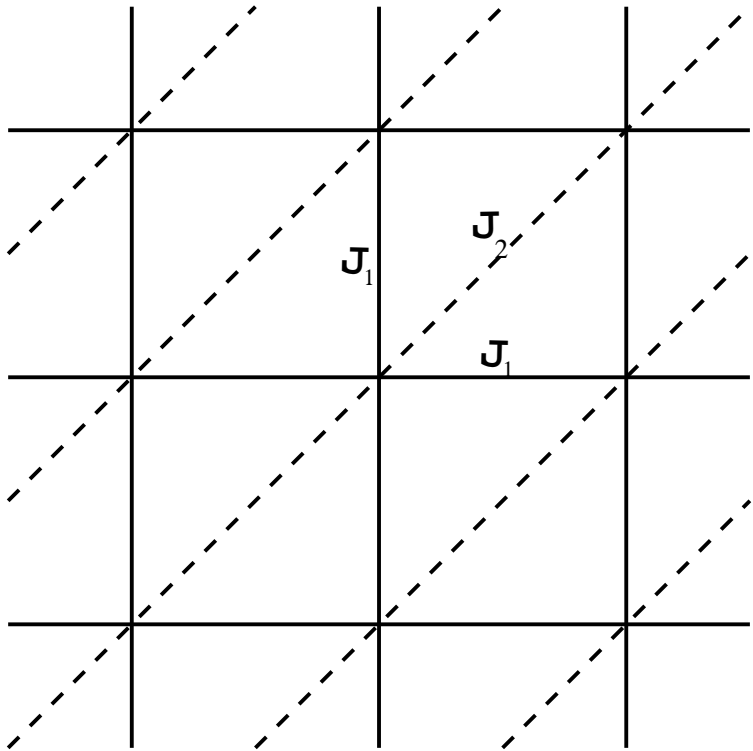}
\caption{
The anisotropic triangular lattice, showing the competing interactions
$J_1$ and $J_2$ that lead to magnetic frustration.
}
\label{lattice}
\end{figure}

We consider the Hubbard model on the anisotropic triangular lattice
with one electron per site. 
If the Coulomb repulsion between two electrons on the same site is sufficiently
large then the ground state is an insulator and a standard strong-coupling 
expansion for the Hubbard Hamiltonian implies that the spin degrees of 
freedom are described by a spin-${{1}\over{2}}$ Heisenberg model
\begin{equation}
H=J_1\sum_{\langle ij \rangle} {\bf S}_i \cdot {\bf S}_j+
J_2\sum_{\langle lm\rangle}{\bf S}_l \cdot {\bf S}_m 
\label{QHAF}
\end{equation}
We use the notation $\langle ij \rangle$ to denote 
nearest-neighbours bonds and $\langle l m \rangle$ to denote bonds along the 
north-east diagonals. $J_1$ is an antiferromagnetic exchange between nearest 
neighbours, {\it i.e.}, along the vertical and horizontal directions and $J_2$ 
is an antiferromagnetic exchange along {\it one} of the diagonals.
$J_1$ and $J_2$ are competing
interactions leading to magnetic frustation. The model is equivalent to a
Heisenberg model on an anisotropic triangular lattice. This lattice is 
shown in Fig.\ref{lattice}.

Note that the special cases $J_2=0$,~$J_1=J_2$, and $J_1=0$ correspond to the
square lattice, isotropic triangular lattice and decoupled chains, respectively. 
 The parameter 
values in Reference \onlinecite{McKenzie:98} suggest that 
${{J_2}\over{J_1}} \sim$ 0.3-1 in the $\kappa-(BEDT-TTF)_2X$ family and so
magnetic frustation will play an important role in these materials.  

Insight can be gained by considering various limits of this model:

{\it Classical limit}. The limit of
infinite spin ($S \rightarrow \infty$) corresponds to a classical Heisenberg model.
The classical ground state of the system can be computed straight-forwardly as
a function of the ratio $J_2/J_1$ assuming that the set of possible
spin configurations of the system are correctly described by a spiral form: 
$ S_i = S {\bf u} e^{i {\bf q} {\bf r}_i } $. {\bf u} is a vector 
expressed in terms of an arbitrary orthonormal basis and {\bf q} defines the
relative orientation of the spins on the lattice. This ground
state configuration was firstly analyzed in \cite{Villain:59}.

Introducing $S_i$ in hamiltonian (\ref{QHAF}), we get an expression for
the classical energy in terms of the spiral vector ${\bf q}$ 
\begin{equation}
{{E(q_x, q_y)}\over{NS^2}} = J_1[\cos q_x+\cos q_y]
+ J_2\cos(q_x+q_y) \equiv  J({\bf q})
\label{EN}
\end{equation}
where $N$ is the number of sites on the lattice. 
For $J_2<J_1/2$, the ground state has N\'eel order with associated wave vector
$\bf{q}$=($\pi, \pi$). For $J_2 > J_1/2$, the ground state has spiral order with
wave vector $(q,q)$ where $q=\arccos(-J_1/2J_2)$.
 
{\it Limiting cases for S=${{1}\over{2}}$.} If $J_2=0$ then the model
reduces to the Heisenberg model on a square lattice.
At zero temperature there will be long range N\'eel order with magnetization 
\cite{Auerbach} of $\langle S_i^z \rangle = 0.3$. If $J_1$ is non-zero but small
it will introduce
a small amount of magnetic frustation which will reduce the magnitude of the 
magnetization in the N\'eel state.
If $J_2=J_1$ then the model reduces to the Heisenberg model on an isotropic
triangular lattice.
There has been some controversy about the ground state of this model. Anderson
\cite{Anderson:73}
originally suggested that the ground state was a "spin liquid" with no 
long-range magnetic order. However, recent numerical work suggests that there
is long-range order but the quantum fluctuations 
are so large due to magnetic frustation that the magnetic moment may be an order
of magnitude smaller than the classical value\cite{Elstner:93}.
 If $J_1$ is non-zero but small we have chains  on the diagonals of the lattice
that are weakly coupled. The case of only two chains corresponds to the "zig-zag" 
spin chain which is equivalent to a single chain with nearest-neighbour and 
next-nearest neighbour exchange, $J_1$ and $J_2$, respectively.
This spin chain has been 
extensively studied and is well understood\cite{Bursill:95}. 
In the limit of interest here,
$J_2 \gg J_1$, there is a gap in the spectrum $\Delta \sim \exp(-const.J_2/J_1)$ and
there is long-range dimer order and incommensurate spin correlations. McKenzie
speculated that this "spin-gap" is still present in the many chain limit 
\cite{McKenzie:98}.  
 
\section{Linear spin-wave theory}
\label{sec:theo}
Extensive theoretical work has been done with the aim of achieving an 
understanding of the ground state
properties of the Heisenberg model on the square and isotropic triangular
lattices. One standard and simple way used for calculating the magnetization
and energy of the magnetically ordered phases of these systems, is linear
spin-wave theory (LSW). LSW has satisfactorily reproduced the ground state
energy and magnetization of the square\cite{Anderson:52,Auerbach} and triangular
lattices\cite{Jolicoeur:89,Jolicoeur:90}. More sophisticated
methods, such as the variational approach of Huse and Elser\cite{Huse:88} 
have corroborated this fact. In the
present work, we apply linear spin-wave theory to the model of interest.
Recently, Bhaumik and Bose \cite{Bhaumik:98} considered the linear spin 
wave theory for the N\'eel phase ($J_2/J_1 < 0.5$) of the same model. 
However, they did not evaluate the quantum corrections to the magnetization. 

Following Miyake\cite{Miyake:91} and Singh and Huse\cite{Singh:92}, 
it is convenient to rotate the quantum
projection axis of the spins at each site along its classical direction.
This transformation is done by introducing the following rotated spin operators
\begin{eqnarray}
S_i^x &=& \hat{S_i}^x\cos(\theta_i)+\hat{S_i}^z\sin(\theta_i) 
\nonumber \\
S_i^y &=& \hat{S_i}^y
\nonumber \\
S_i^z &=& \hat{S_i}^z\cos(\theta_i)-\hat{S_i}^x\sin(\theta_i) 
\nonumber \\
\label{ROT}
\end{eqnarray}
in Hamiltonian (\ref{QHAF}).

This rotation simplifies the spin-wave treatment so that only one type of bosons
rather than three is needed to describe the spin operators. After this
transformation the Hamiltonian is
\begin{eqnarray}
H &=&J_1\sum_{\langle ij \rangle}{\cos (\theta_i-\theta_j)(\hat{S}_i^x\hat{S}_j^x-
\hat{S}_i^z\hat{S}_j^z)+\sin
(\theta_i-\theta_j)(\hat{S}_i^z\hat{S}_j^x-\hat{S}_i^x\hat{S}_j^z)+\hat{S}_i^y
\hat{S}_j^y}  
\nonumber \\
&+&J_2\sum_{\langle lm \rangle}{\cos (\theta_l-\theta_m)(\hat{S}_l^x\hat{S}_m^x
-\hat{S}_l^z\hat{S}_m^z)+\sin(\theta_l-\theta_m)(\hat{S}_l^z\hat{S}_m^x-\hat{S}_l^x
\hat{S}_m^z)+\hat{S}_l^y\hat{S}_m^y}-B\sum_i{\hat{S}_i^z}  
\nonumber \\
\label{HAF}
\end{eqnarray}
where we have introduced an auxiliary magnetic field, $B$, in the $z$ direction of 
rotated basis in order to compute the magnetization.   

Using the standard Holstein-Primakoff representation\cite{Auerbach} for the
spin operators we expand them with respect to $1/S$ and we take the Fourier 
Transform of the boson operators. The resulting Hamiltonian  
up to order $O(1/S)$
\begin{eqnarray}
H &=& J_1S^2\sum_{\langle ij \rangle} \cos(\theta_i-\theta_j) + 
J_2 S^2 \sum_{\langle lm\rangle} \cos(\theta_l-\theta_m) - NBS
\nonumber \\
&+&S\sum_{\bf k} (J_1A_1({\bf k})+J_2A_2({\bf k}) -J_1 C_1 
-J_2C_2 +B/S)a^+_{{\bf k}}a_{{\bf k}} 
\nonumber \\
&+& S \sum_{\bf k}(J_1B_1({\bf k})+ J_2B_2({\bf k}))(a^+_{\bf k}a^+_{\bf -k}+a_{\bf
k}a_{\bf -k})
\nonumber \\
\label{KSP}
\end{eqnarray}
is not diagonal in the boson operators. 
$N$ is the number of lattice sites in the system and the coefficients $A$, 
$B$ and $C$ are 

\begin{eqnarray}
A_1({\bf k}) &=& {{1}\over{N}}\sum_{\langle ij\rangle} \cos({\bf k}({\bf r}_j-{\bf r}_i) ) 
(\cos(\theta_i-\theta_j) +1) 
\nonumber \\
A_2({\bf k}) &=& {{1}\over{N}}\sum_{\langle lm \rangle} \cos({\bf k}({\bf r}_m-
{\bf r}_l) ) (\cos(\theta_l-\theta_m) +1)
\nonumber \\
B_1({\bf k}) &=& {{1}\over{2N}}\sum_{\langle ij \rangle} \exp(i{\bf k}({\bf r}_j-
{\bf r}_i) ) (\cos(\theta_i-\theta_j) -1)
\nonumber \\
B_2({\bf k}) &=& {{1}\over{2N}}\sum_{\langle lm \rangle} \exp(i{\bf k}({\bf r}_m-
{\bf r}_l ) ) (\cos(\theta_l-\theta_m) -1)
\nonumber \\
C_1 &=& {{2}\over{N}} \sum_{\langle ij \rangle} \cos(\theta_i-\theta_j)
\nonumber \\
C_2 &=& {{2}\over{N}} \sum_{\langle lm \rangle} \cos(\theta_l-\theta_m)
\nonumber \\
\end{eqnarray}
where again $\langle ij \rangle$ sums over the nearest neighbours and
$ \langle lm \rangle $ over the next-nearest ones. 

Hamiltonian (\ref{KSP}) explicitly breaks up the SU(2) symmetry of the spins at each
site. Only magnetically ordered states can be
analyzed with this method and the expansion is valid only when the correction 
due to the zero-point motion of the spins is sufficiently small.
 
Following standard LSW theory we   
diagonalize hamiltonian (\ref{KSP}) using a Bogoliubov transformation, 
assuming, as we have already discussed, an spiral ordering of the spins
in the lattice, so that the relative angle between two
nearest-neighbour spins is $q$, and $2q$ for two next-nearest-neighbours.
The diagonal Hamiltonian reads                 
\begin{eqnarray}
H &=& NS^2 J({\bf q})-NBS                        
\nonumber \\
&-&{\frac{{1}}{{2}}}S\sum_{{\bf k}}\{ {{1}\over{2}}( J({\bf k}+{\bf q}) + J({\bf k}) ) 
+ B/S - 2 J({\bf q}) - \omega({\bf k},B)\}
\nonumber \\
&+&\sum_{{\bf k}}\omega({\bf k,}B)\alpha _{{\bf k}}^{+}\alpha _{{\bf k}} 
\label{Hamilt}
\end{eqnarray}
where the sums run over the first Brioullin zone and $\alpha_k$ creates 
boson spin excitations (magnons) with the dispersion relation
\begin{equation}
\omega({\bf k,} B) = 2 S [ (J({\bf k})+ B/S -J({\bf q}) )( {{1}\over{2}} 
(J( {\bf k}+ {\bf q}) + J( {\bf k} - {\bf q} ) ) + B/S - J({\bf q}) ) ]^{1/2} 
\label{POLES}
\end{equation}
Equation (\ref{POLES}) explicitly shows the zero bosonic modes  
at the momentum wavevectors, {\bf k}={\bf q} and {\bf k}=0, for $B=0$. 

The ground state energy of the system, $E_0$, is given by equation (\ref
{Hamilt}), setting the occupation of the bosons to zero, and,  
finally, the magnetization of the system is computed as the derivative of
the ground state energy: 
\begin{eqnarray}
&<&S_i^z>=-{\frac{{1}}{{N}}}\lim_{B->0}{\frac{{dE_0}}{{dB}}}  
\nonumber \\
&=&S+{{1}\over{2}}-  S{{1}\over{2N}}\sum_{{\bf k}} { { {{1}\over{2}} ( J( {\bf k}
+ {\bf q} ) + J( {\bf k} -{\bf q} ) ) + J( {\bf k} ) - 2 J( {\bf q} ) } \over{
\omega({\bf k},B=0) } }  
\label{spin}
\end{eqnarray}

This expression recovers the LSW expression for the magnetization
on the square lattice\cite{Auerbach}, {\bf q}=($\pi,\pi $), and the isotropic 
triangular lattice\cite{Miyake:91}, ${\bf q}=(2\pi/3, 2\pi/3)$.

\section{Results}
\label{sec:res}  
We have numerically evaluated the integrals in equation (\ref{Hamilt})
and (\ref{spin}) to obtain the ground state energy and magnetization 
as a function of $J_2/J_1$.                                                     

\subsection{Ground state energy}

\begin{figure}
\center
\epsfxsize = 10cm \epsfbox{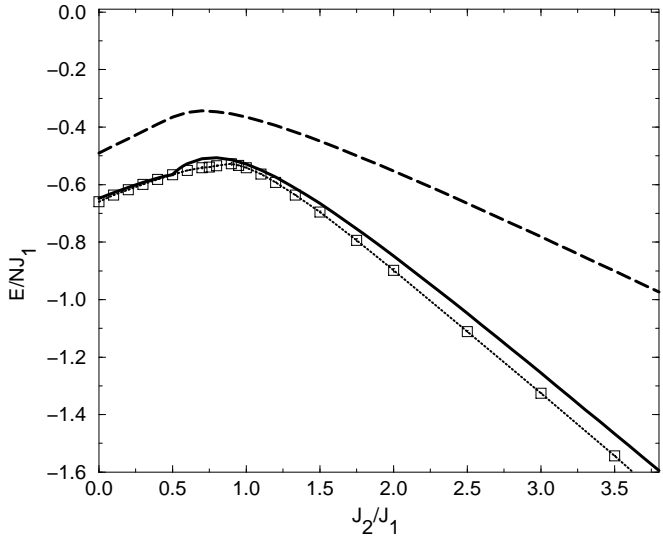}
\caption{
The total ground state energy (full line) for $S={{1}\over{2}}$ and classical
ground state energy (dashed line) as a function of the amount of magnetic
frustration $J_2/J_1$. The squares represent the ground
state energies obtained from a series expansion calculation\cite{Zheng:98}.
The error bars associated to the series expansion energies are, at
most, of order 10$^{-3}$, so they cannot be seen in this plot.
}
\label{energy}
\end{figure}

In Fig.~\ref{energy} we plot the ground state energy per site as a 
function of $J_2/J_1$.
The classical
energy is also included in the same figure (dashed line),
showing how quantum fluctuations lower the ground state energy. The maximum in 
the total energy is attained 
around $J_2/J_1 \approx 0.7$, which approximately coincides with the position of the 
maximum in the classical energy at $\sqrt{2}/2$: at this point, geometrical 
frustration attains its maximum. A cusp is found at the transition from the
N\'eel to the spiral phase at $J_2/J_1 = 0.5$, which
results, as will be later seen, from the softening of the spin-wave modes
at the transition point. We also plot in the same figure, results for
the ground state energy obtained from a series expansion calculation\cite{Zheng:98}. 
Although LSW is a simple approximation for computing ground state properties
of frustrated systems, the energies obtained are in 
good agreement with the series expansion results. 

\subsection{Magnetization}
\begin{figure}
\center
\epsfxsize = 10cm \epsfbox{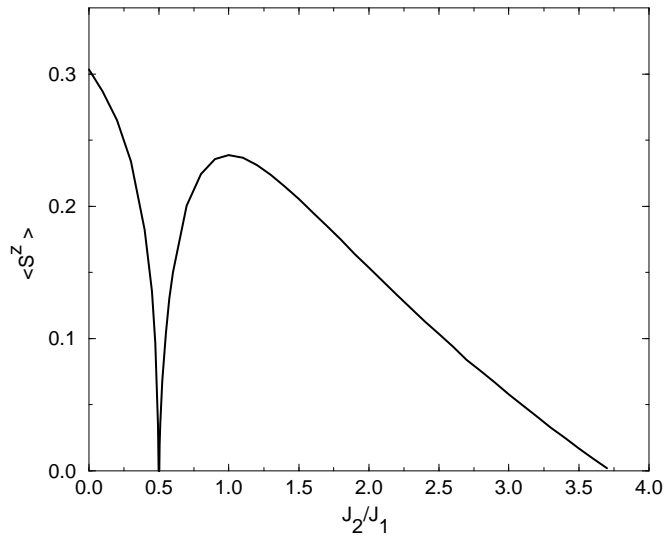}
\caption{
The effect of frustration on the magnetization calculated from linear spin-wave
theory is shown as a function of $J_2/J_1$. The quantum correction to the magnetization
diverges as $J_2/J_1 \rightarrow \infty$. At the N\'eel-spiral
transition (at $J_2/J_1=0.5$)
the correction is large due to the softening of the
spin-wave modes but finite (see the text). This suggests the
possibility of a quantum disordered phase
at $J_2/J_1 \approx 0.5$ and for $J_2/J_1 > $ 4. }
\label{magnet}
\end{figure}

In Fig.~\ref{magnet} we show results for the magnetization for
$S={{1}\over{2}}$ as a 
function of $J_2/J_1$. The results are qualitatively similar to those
of a recent series expansion study \cite{Zheng:98}. We have tested 
the accuracy of the calculation 
by comparing with known results for the square
($J_2/J_1=0.0$) lattice \cite{Auerbach} with magnetization $ \langle S_i^z \rangle =
0.30339 $, and the triangular lattice \cite{Jolicoeur:89} ($J_2/J_1$=1.0) with
$ \langle S_i^z \rangle = 0.23868 $.
We find a strong dip in the magnetization at $J_2/J_1=0.5$ suggesting the 
possibility of a disordered phase in its neighbourhood.              
The nature of the ground state is unclear and will have to be determined
by more sophisticated techniques.  

It is instructive to mention results reported on other lattices
such as the $J_1-J_2$ square lattice (for which there is frustration along 
both diagonals), for which extensive work has been 
carried out.  
Schulz, Ziman and Poilblanc\cite{Schulz:96}, using exact diagonalization
of finite cells, get results for the magnetization 
in qualitative agreement with LSW theory and a Dyson-Maleev approach
performed by Gochev\cite{Gochev:94}
which treats the interaction between the spin-waves self-consistently. 
However, on the isotropic triangular lattice with next-nearest neighbours,
Deutscher and Everts ~\cite{Deutscher:93} have shown that, while   
exact diagonalization shows a finite jump of the magnetization between
the collinear and the $q=2\pi/3$ canted phases, LSW theory gives a
continuous transition.    

From the above discussion 
we therefore conclude that it is difficult to extract conclusive answers     
from LSW theory for this
frustated lattice at the transition point: the interaction between the 
spin-waves becomes very large due to the presence of geometrical 
frustation leading to the possibility of a completely 
different state than the classical one and more sophisticated methods are 
needed to describe this region correctly. A recent series expansion 
study \cite{Zheng:98} suggests that the system is N\'eel ordered for
$J_2/J_1 < 0.7$ and quantum disordered for $0.7 < J_2/J_1 < 0.9$. 

\section{Spin-wave velocities and quantum fluctuations}

\subsection{Near N\'eel-spiral transition}
\begin{figure}
\center
\epsfxsize = 10cm \epsfbox{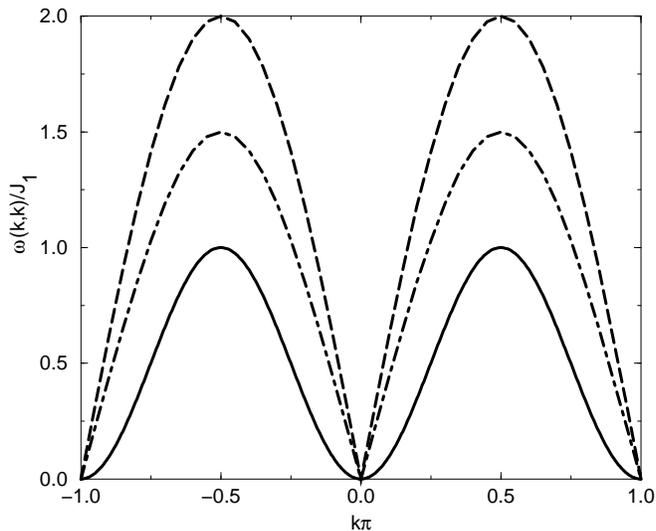}
\caption{
The softening of the spin-waves near the commensurate-incommensurate transition.
The spin-wave dispersion $\omega(k_x, k_y)$ along the $(k,k)$ direction
is shown for different values of $J_2/J_1$. Note that the dispersion
near 0 and $\pi$ becomes quadratic at the transition. As discussed in
the text, this leads to an increase in the quantum correction
to the magnetization. $J_2/J_1=0$ (dashed line), $J_2/J_1=0.25$ (dashed-
dotted line) and $J_2/J_1=0.5$ (full line).
}
\label{magnon1}
\end{figure}

Insight into the origin of the increase in the quantum fluctuations
near $J_2/J_1$=0.5, can be gained from considering 
the behaviour of the
spin-wave velocities near the zero bosonic modes: (0,0) and $(\pi, \pi)$. 
In Fig.~\ref{magnon1}, we show the dependence of the magnon frequencies
as a function of the wavevector $k$ along the $(k, k)$ direction in
the parameter region $ 0 < J_2/J_1 \leq 0.5 $ where the $(\pi, \pi)$
state is classically stable. Except at $J_2/J_1 = 0.5$ the magnon
excitation vanishes linearly with wavevector and we define the 
associatied spin-wave velocity along the
$(k, k)$ direction as $c_+$ and $c_-$ along the direction perpendicular 
to it: $(k, -k)$. 

While near the zero modes, (0,0) and $(\pi,\pi)$,
the spin-wave velocity $c_+ \approx 2 S J_1 \sqrt{ 2 ( 1- 2 J_2/J_1)}$, 
$c_-$ does not depend on $J_2$, $c_- \approx 2 \sqrt{2} S J_1$. Hence,
the modes soften along the
diagonal direction and the velocity $c_+$ vanishes at the N\'eel-spiral 
transition. This can be clearly seen in Fig.~\ref{magnon1}.

From the above analysis we can now explain  
the behaviour of the low-energy, long-wavelength component of 
the quantum-fluctuations
as $J_2/J_1$ approaches 0.5. At this point and near $(k_x, k_y)
\approx (0,0)$, the correction to the magnetization due to quantum fluctuations
(see Eqn.(\ref{spin})) for $S={{1}\over{2}}$ can be approximated by
\begin{eqnarray}
&\langle& S^z_i \rangle - {{1}\over{2}} \approx {{1}\over{2}} - 
{{1}\over{8 \pi^2}} \int{ {{dk_+ dk_-} \over{ \sqrt{ ({{k_+}\over{2}})^4 + 
k_-^2 }}}}  
\nonumber \\
&\approx& {{1}\over{2}} - {{k_c}\over{2 \pi^2}}(\log(2k_c) - 2 \log(k_c/2) +2)
\nonumber \\
\label{correc}
\end{eqnarray}
where the integral is expressed in terms of the diagonal directions,
$k_+=k_x+k_y$ and $k_-=k_x-k_y$, and $k_c$ is a cutoff in the momentum. 
As expected, the lowest order term that appears in the integral 
along the $k_+$ direction is quartic, because the spin-wave velocity vanishes, 
$c_+ \rightarrow 0$. This behaviour of the magnon spectrum, where
the softening occurs only along {\it one} direction, has also been found
by Chubukov and Jolicoeur\cite{Chubukov:92} on the isotropic triangular lattice
with next-nearest neighbours at the transition from the collinear to the
incommensurate
phase. They were
able to show that the transition point shifts from the LSW theory
result when treating the effect of
quantum fluctuations in a self-consistent way. This is in agreement with
the results obtained using series
expansions \cite{Zheng:98}, which give the transition point at $J_2/J_1$=0.7,
instead of 0.5 obtained within LSW theory.
   
As shown in (\ref{correc}) the integral is {\it finite} but 
gives a much larger correction than for the square lattice $(J_2=0)$ case.
We find that taking a cutoff of $0.1 \pi$, 
the integral in the correction is two to three times larger than 
for the square lattice ($ J_2=0 $), for the same region of integration. 
Therefore, the reduction in energy of the bosonic modes along 
the $k_+$ direction (see Fig. \ref{magnon1}) 
is responsible for a finite but
large enhancement of the quantum correction to the magnetization.   

It is interesting to compare this result with the one obtained on 
the $J_1-J_2$ square lattice: in this case, the correction due to the  
quantum fluctuations at $J_2/J_1 = 0.5$, can be approximated by
\begin{eqnarray}
&\langle& S^z_i \rangle - {{1}\over{2}}\approx {{1}\over{2}} -
{{1}\over{8 \pi^2}} \int{ {{ dk_+ dk_-}
\over{ \sqrt{ ({{k_+}\over{2}})^4 +
({{k_-}\over{2}})^4 - {{k_+^2 k_-^2}\over{8}} }}}}
\nonumber \\
&=& {{1}\over{2}} - {{1}\over{(2\pi)^2}} 
\int{ {{dk_x dk_y}\over{k_x k_y}} }
\nonumber \\
\label{J1J2}
\end{eqnarray}
the last integral in Eqn.(\ref{J1J2}) shows a divergence that
behaves like the square of a logarithm.
Therefore, we find a different qualitative behaviour of the
quantum fluctuations for the anisotropic lattice   
as compared to the $J_1-J_2$ square lattice.  

\subsection{Weakly coupled chains ($J_2 >> J_1$)}
\begin{figure}
\center
\epsfxsize = 10cm \epsfbox{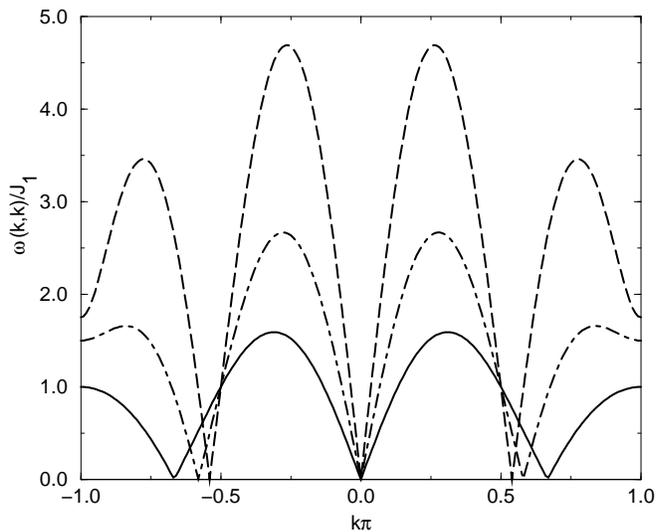}
\caption{
The spin-wave dispersion relation along the $(k,k)$ direction
in the spiral phase. As $J_2/J_1$ increases the limit of weakly
coupled chains is approached and the classical ordering wavevector, {\bf q},
at which the spin-wave energy vanishes, approaches $\pi/2$.
$J_2/J_1=1$ (full curve), $J_2/J_1=2$ (dashed curve) and
$J_2/J_1 =4$ (dashed-dotted curve).
}
\label{magnon2}
\end{figure}

We have analyzed the structure of the magnon excitation spectra 
in the neighbourhood of the zero bosonic eigenenergies: 
$(0,0)$ and $(q,q)$, with $q \rightarrow \pi/2$  as $J_2/J_1 \rightarrow \infty$. 
We find that the spin-wave velocities are $c_-\approx \sqrt{2} S J_1$ 
and $c_+\approx 2 \sqrt{2} S J_2$. 
The dispersion relation is plotted in Fig.~\ref{magnon2}.
Again, we analyze the low-energy, long-wavelength contributions to
the integral appearing in Eq.~(\ref{spin}). Near the zero modes: $(0,0)$,
and  $(q,q)$, it can be approximated by an elliptic integral  
\begin{equation} 
\int{ {{dk_+ dk_-}\over{ \sqrt{ (c_+ k_+)^2 + (c_- k_-)^2 } } }  }
\end{equation}
In the limit $J_2/J_1 >>1$, the ratio of velocities diverges 
as $c_+/c_- \approx 2 J_2/J_1 $, and the elliptic integral can be    
approximated by a logarithm. Taking a cut-off of $\pi/2$,
we get for the magnetization (Eq.\ref{spin})
\begin{equation}
<S_i^z> \approx S + 3 {{\sqrt{2}}\over{4\pi}} \log(J_1/2 J_2)
\end{equation}
As expected, the quantum fluctuations diverge as $J_2/J_1 \rightarrow \infty$. 
For $S={{1}\over{2}}$ the critical value that sets $<S_i^z>= 0$ is  $J_2/J_1 \approx 
2$. Although this is just a crude estimate it is roughly consistent with Fig.
\ref{magnet}. 

It is interesting to compare this result with 
the one obtained by Affleck, Gelfand and Singh\cite{Affleck:94} for the 
anisotropic $J_y-J_x$ square lattice where $J_y$ is a non-frustrating
interaction which can be gradually turned off. In the limit 
$J_y/J_x \rightarrow 0$, the integral reduces to 
\begin{equation}
<S_i^z> \approx S + {{1}\over{2 \pi}} \log(J_y/J_x)
\end{equation}
taking a momentum cut-off of $\pi$. Note that the factor 
multiplying the logarithm is smaller in this case than in
the anisotropic triangular lattice, and, therefore, 
the critical value obtained, within spin-wave theory, is 
an order of magnitude larger: $J_x/J_y \approx 23.1$. 
Numerical work\cite{Affleck:94} and renormalization group 
arguments \cite{Affleck:96,Wang:97} on the 
anisotropic square lattice suggests the existence
of long range N\'eel order for an infinitesimal coupling between 
the chains ($J_y/J_x \ll 1$). A similar renormalization group
analysis for the appropiate SO(3) nonlinear sigma model 
\cite{Azaria:92,Chubukov:94}, could be performed to gain some insight into
the ground state of our model in this parameter region. Also numerical calculations
using more sophisticated numerical techniques could be performed 
to find whether long range order persists for an infinitesimal
coupling or not.  
 However, our analysis clearly shows that 
for comparable interchain coupling the quantum fluctuations are
much larger than for the case where the interchain coupling is
non-frustrated.   

\section{Conclusions}
\label{sec:concl}

 We have presented a linear spin-wave analysis of the Heisenberg
antiferromagnetic Heisenberg model on an 
anisotropic triangular lattice. The Heisenberg model on this
lattice should be the relevant model for 
describing the insulating properties of certain layered organic superconductors. 
The correction to both, the energy and magnetization due to quantum fluctuations,
is computed by means of linear spin-wave theory for different values of $J_2/J_1$.
The results obtained from LSW theory suggest the possibility of finding 
a disordered state near $J_2/J_1 \approx 0.5$. This possibility also
exists for values of $J_2/J_1 > 4$, where the system resembles a set
of chains weakly coupled by a frustrated interaction. In this
region of parameters we find that quantum fluctuations are larger than
in the case where the chains are weakly coupled by a non-frustrated
interaction.
Further work using other numerical approaches should be used to analyze
in more detail the results presented here. 
\acknowledgements
We thank J. Oitmaa and R. R. P. Singh for helpful discussions.  
We thank W. Zheng for providing us with the series expansion data. 
Computational work in support of this research was partly performed on 
Cray computers at Brown University's Theoretical Physics Computing Facility.
J. Merino was supported in part by a NATO postdoctoral fellowship  
and the Australian Research Council. This work was supported 
in part by the National Science Foundation under contracts NSF DMR-9357613
(J. B. M. and C. H. C.).  
\vskip 1.0 cm 
Note added: When we were completing this manuscript we became aware
that Trumper \cite{Trumper:98} had obtained some of the results presented
here.

\end{document}